\documentstyle [12pt] {article}
\title{QUASI-CLASSICAL DYNAMICAL DETERMINATION OF THE BASIC PLANCK UNITS}

\author{Vladan Pankovi\'c, Darko V. Kapor, Miodrag Krmar\\
Department of Physics, Faculty of Sciences, 21000 Novi Sad,\\ Trg
Dositeja Obradovi\'ca 4., Serbia, vpankovic@if.ns.ac.yu}

\date {}
\begin{document}
\maketitle \vspace {0.5cm}
 PACS number:  03.65.Ta
 \vspace {0.5cm}

\begin {abstract}
In this work all basic Planck units (Planck mass, Planck length,
Planck time and Planck electrical charge) are determined by
solution of a simple system of quasi-classical dynamical equations
that have particularly classical (Newton gravitational  force,
Coulomb electrostatic force), relativistic (equivalence principle)
or quantum (Planck-Einstein formula) nature.
\end {abstract}

\vspace {0.5cm}

As it is well-known Plank gave his remarkable system of the basic
units, Planck mass, Planck length, Planck time and Planck
electrical charge, without any deduction rule (algebraic formulas
referring to some physical, i.e. dynamical principles) [1].
Practically, there is common opinion, that deduction of the basic
Planck unit in low energetic domain, i.e. non-Planckian sector,
represents nothing more than dimensional analysis only [2].
Situation would be quite different in high energetic, Planckian
domain where new, to this time undone, quantum field theoretical
and string dynamical laws have primary role [3].

In this work all basic Planck units (Planck mass, Planck length,
Planck time and Planck electrical charge) will be determined by
solution of a simple system of quasi-classical dynamical equations
that have particularly classical (Newton gravitational  force,
Coulomb electrostatic force), relativistic (equivalence principle)
or quantum (Planck-Einstein formula) nature.

Consider the following system of four equations
\begin {equation}
        \frac {Gm^{2}}{R^{2}} = \frac {1}{4 \pi \epsilon_{0}}\frac {e^{2}}{R^{2}}
\end {equation}
\begin {equation}
        \frac {Gm^{2}}{R^{2}} = m \omega^{2}R
\end {equation}
\begin {equation}
        \frac {Gm^{2}}{R^{2}} = mc^{2}
\end {equation}
\begin {equation}
        mc^{2} = \hbar \omega              .
\end {equation}
Here m represents the mass of a system, e- electrical charge of a
system, R - distance between two systems, and, $\omega$ - angular
frequency, all of which can be considered as four unknown
variables. Also, in (1)-(4) G represents Newtonian gravitational
constant, $\epsilon_{0}$ - vacuum dielectric constant or
permittivity, c - speed of light and $\hbar$ - Planck reduced
constant.

Equation (1) can be, formally, interpreted as the dynamical
equilibrium between classical Newton gravitational force (that
originates from one particle with mass m and that acts on the
other particle with the same mass m) and classical Coulomb
electrostatic force (that originates from one particle with
electrical charge e and that acts on the second particle with the
same electrical charge e). It, in some degree corresponds to Stone
concept by derivation of the basic units, especially basic mass
[4].

Equation (2) can be, formally, interpreted as the dynamical
equilibrium between mentioned classical Newton gravitational force
and classical centrifugal force by rotation of the particle. It
can be observed that instead of centrifugal force an elastic force
of a linear harmonic oscillator or rotator can be used.

In this way both equations, (1) and (2), have, formally, a
classical dynamical nature.

Equation (3) can be, formally, interpreted as the equivalence
between potential energy of mentioned classical gravitational
interaction and relativistic total energy of single particle. In
this way given equation has dynamically a half-classical and
half-relativistic nature.

Finally, equation (4) can be, formally, interpreted as the
quantum, Einstein-Planck relation for energy of a quantum. In this
way given equation has, formally, a quantum dynamical form.

Physical, precisely quasi-classical, meaning of the complete
system of equations (1)-(4) is the following. Basic Planck units
are quasi-classically determined by conditions that both classical
force, gravitational and electrical, between two equivalent
systems, are identically strong, as well as by
quantum-relativistic condition that any of interacting system
becomes comparable, even equivalent with quantum of given
interactions. (Emission and absorbtion of interaction quantum can
be, in the first approximation, considered as the harmonic
oscillating.) It, of course, cannot represent a complete, exact
meaning of basic Planck unit, but it represents a satisfactory,
effective quasi-classical interpretation. As it is not hard to see
such interpretation refers to future exact interpretation
similarly to Bohr atomic theory, i.e. naïve quantum theory of atom
to exact, quantum mechanical theory of atoms.

Now we shall solve system of equations (1)-(4).

According to (1) it follows
\begin {equation}
        e =  (G4\pi \epsilon_{0})^{\frac {1}{2}}m               .
\end {equation}

According to (3) it follows
\begin {equation}
        R = \frac {Gm}{c^{2}}                       .
\end {equation}

According to (4) it follows
\begin {equation}
        \omega = \frac {mc^{2}}{\hbar}                         .
\end {equation}

Introduction of (5)-(7) in (2), after simple calculation, yields,
\begin {equation}
       m = (\frac {\hbar c^{3}}{G})^{\frac {1}{2}}                    .
\end {equation}

Now, introduction of (8) in (5)-(7), after simple calculation,
yields
\begin {equation}
         e =  (\hbar c 4 \pi \epsilon_{0}) ^{\frac {1}{2}}
\end {equation}
\begin {equation}
       R = (\frac {\hbar G}{c^{3}})^{\frac {1}{2}}
\end {equation}
\begin {equation}
      \omega = (\frac {c^{5}}{\hbar G})^{\frac {1}{2}}                    .
\end {equation}

In this way unique, simple solution of the system of equations
(1)-(4) is obtained.

As it is not hard to see m (8) is identical to Planck mass, e (9)
- to Planck electrical charge, R (10) - to Planck length, while
$\omega$ (11) is identical to inverse , i.e. reciprocal value of
Planck time.

In conclusion it can be repeated and pointed out the following. In
this work all basic Planck units (Planck mass, Planck length,
Planck time and Planck electrical charge) are determined by
solution of a simple system of quasi-classical dynamical equations
that have particularly classical (Newton gravitational  force,
Coulomb electrostatic force), relativistic (equivalence principle)
or quantum (Planck-Einstein formula) nature.

\vspace{0.5cm}

 {\large \bf References}

\begin {itemize}

\item [[1]] M. Planck, S.-B. Preuss. Akad. Wiss. {\bf 5} (1899) 440
\item [[2]] P. W. Bridgman, {\it Dimensional Analysis} (Yale University Press, New Haven - London, 1931)
\item [[3]] M. Duff, L. B. Okun, G. Veneziano, Journal of High Energy Physics {\bf 3}  (2002) 23 ; class-phys/0110060 and references therein
\item [[4]]  J. Stoney, Phil. Trans. Roy. Soc. {\bf 11}  (1881) 381

\end {itemize}

\end {document}